Francois-Igor PRIS (И. Е. Прись)

PhD in theoretical physics, PhD in philosophy
Institute of philosophy of the National Academy of Sciences of Belarus
frigpr@gmail.com

# ON "AGENTIAL REALISM" AND ONTOLOGY OF QUANTUM MECHANICS[1]

*Abstract.* K. Barad proposes "agential realism" as a unified approach to natural and social phenomena. The position is inspired by quantum mechanics and in particular the phenomenon of quantum entanglement. Barad also sees similarities between her approach, N. Bohr's view and C. Rovelli's relational quantum mechanics. In our view, agential realism is a kind of ontological correlationism, not a realism. The analogy with the Bohr and Rovelli approaches is only partial. Agential realism is a wrong interpretation of quantum mechanics. It is also unsuitable for social theorizing, for which taking into account the sensitivity of ontology to the context is fundamental. As an alternative, we propose a contextual quantum realism (CQR) that rejects substantive dualisms (as does Barad), but at the same time accepts the categorical dualism of the real and the ideal. Our approach also allows one to better understand Bohr's position and to correct Rovelli's relational quantum mechanics.

Karen Barad has proposed a unified approach to natural and social phenomena – "agential realism", which she believes can be derived from quantum mechanics and represents a development of N. Bohr's position [1–2]. Barad writes: "'Position' only has meaning when a rigid apparatus with fixed parts is used (e.g., a ruler is nailed to a fixed table in the laboratory, thereby establishing a fixed frame of reference for specifying 'position'). And furthermore, any measurement of 'position' using this apparatus cannot be attributed to some abstract independently existing 'object' but rather is a property of the phenomenon – the inseparability of 'observed object' and 'agencies of observation'" [1, p. 814]. The first half of this quotation is consistent with our position which asserts the sensitivity of ontology to the context and which we call "contextual quantum realism" (CQR), while the second half, beginning with "but rather is a property of the phenomenon", contradicts it and can be understood as the position of an ontological correlationism we criticize. Indeed, the American philosopher explains her point as follows: "Phenomena do not merely mark the epistemological inseparability of 'observer' and 'observed'; rather, *phenomena are the ontological inseparability of agentially intra-acting 'components'*. That is, phenomena are ontologically primitive relations – relations without preexisting relata" [1, p. 815]. Thus, ontologically primitive are phenomena as intra-actions between object and instrument of observation, object and agent, not having an independent existence, but being derived from the intra-action – the primary entity. The division into object and measuring instrument, object and quantum subject (agent) is secondary. To back up her position, Barad turns to the notion of quantum entanglement. For her, the interaction of systems (observer and observable) – holistic quantum entanglement – is primary, while the individual components of this holistic system, observer and observable, are secondary. Barad thinks that for Bohr, too, "phenomena" are ontologically primary as relations, inseparable wholes of quantum object and subject. She also says: "According to Bohr (…) our knowledge-making practices, including the use and testing of scientific concepts, are material enactments that contribute to, and are a part of, the phenomena we describe" [2, p. 32]. This is a wrong interpretation of Bohr's position which is not an enactment but a contextualist view.

[1] The reported study was partially funded by BRFFR according to the research project № Г22МС-001 «Quantum realism – contextual realism. (QBism and other interpretations of quantum mechanics from the point of view of a contextual realism.)»

J. Faye notes that "on Barad's reading of Bohr, it is not the quantum object that has a property in an experimental context, but rather it is the relationality of quantum object and experimental set-up, the 'phenomenon', that has the property. This relationality – as captured by the phenomenon – is the cornerstone in Barad's ontological framework and the composition metaphor implicit in the quotation above should only be assigned instructional significance" [3, p. 7]. Faye, a specialist on Bohr, rejects Barad's interpretation of Bohr. For him Bohr's point of view is a kind of relational holism, not ontological, but epistemic [3]. This is also the position of Faye himself. M. Dorato, for example, is an epistemic holist, too [4].

We reject agential realism and try to understand and interpret Bohr's position from the point of view of our CQR. That is, we emphasize that a quantum object has a certain property in an experimental context and only in it. This means that ontology is sensitive to context. At the same time, for example, C. F. von Weizsäcker, M. Bitbol, B. Falkenburg, and some other philosophers consider Bohr's philosophical position as a neo-Kantian one.[2]

Thus, according to agential realism, there are no individual objects with determinate properties which are revealed by measurement. Moreover, this position interprets Bohr's complementarity as a metaphysical indeterminacy of quantum properties before the measurement. From the point of view of our CQR, one should not speak of metaphysical indeterminacy, as if such indeterminacy were an inherent property of reality. The concept of reality does not imply this. Reality as such is neither determinate nor indeterminate; by definition, reality is simply what it is. Wave and corpuscular properties of quantum systems (objects) do not arise as a result of a measurement, within the subject-object correlation, phenomenon. They are literally identified as real properties that exist and existed before their identification. However, before their identification they had no identity because the context was not fixed. That is, there is no point in talking about the determinate physical properties before their identification because the "act" of identification is context-sensitive.

Within her "agential realism", referring to Bohr, Barad questions the dualisms of subject-object, knower-known, culture-nature, word-world [2, p. 147]. Not surprisingly, if one takes the path of erasing the categorical distinction between the ideal and the real, which is simply not perceived as such. Like Barad, our CQR rejects substantial dualisms, but, in contrast, retains the categorical 'dualism' that allows a genuine distinction to be made between realism and anti-realism. For anti-realists, the ideal is real in a particular sense (e.g. metaphysical Platonism) or is part of reality (in this sense, for example, the metaphysical realism of "external world" is a kind of anti-realism because the objects of the world are considered as meaningful in themselves).

According to Faye, Bohr does not want to eliminate the epistemic distinctions between subject and object, knower and known, word and world [3]. He considers the measurement outcome as giving information about an object that exists separated from the measuring instrument. Indeed, Bohr argues for the need to use classical concepts to describe measuring instruments, measurement procedure and the measurement outcomes. By doing so, he assumes that the measurement outcomes and the corresponding properties relate to objects that exist independently from any observer. An essential difference from classical mechanics is that in quantum mechanics classical concepts are applied in the context of experimental observation. In one context, a quantum object exhibits its wave properties, while in another it exhibits its corpuscular properties.[3] Both are real, they do not arise from measurement as properties of "phenomena" in Barad's sense. According to Faye, Bohr does not question epistemic but ontological dualism. The dualisms of subject-object, knower-known, word-world are not questioned. On the contrary, they are at the foundation of his interpretation of quantum mechanics [3].

---

[2] According to Faye, different authors also regard Bohr as an instrumentalist, an objective anti-realist, a phenomenological realist or a realist of various sorts. They, however, use the terms 'realism' and 'anti-realism' differently. Faye holds that Bohr is an objective anti-realist. That is, he is an entity realist and anti-representationalist concerning theories which are seen as symbolic, not literal, representations of reality. For Bohr, atoms and other "unobservable entities" are real; they are not simply heuristic or logical (mathematical) constructions to describe and predict empirical phenomena, but without any physical meaning (instrumentalism). Faye also notes the influence of Kant on Bohr [5]. (According to F. A. Muller, Bohr said that he had not read Kant.)

The categorical distinction between the ideal and the real underlying CQR allows one to better understand Bohr's Copenhagen interpretation, to eliminate the conceptual confusions faced by Barad's agential realism, to correct (contextualise) C. Rovelli's "democratisation of the Copenhagen interpretation", i.e. relational quantum mechanics (RQM) [6–7]. Rovelli keeps all systems, the observer and the observable, on the same footing; he treats the observer as a physical system (he calls it "context") which, when interacting with ("observing") another system, becomes entangled with it. For us, the instrument of observation (observer) belongs to the category of the ideal. Strictly speaking, it does not interact with the observed system. But its status changes, and it begins to belong to the category of the real, if it itself becomes the object of observation [6–7].

Tellingly, Barad believes that her agential realism is also close to Rovelli's RQM [2, p. 333]. As in the case of Bohr, there are some reasons to accept this view. Agential realism is a radical (ontological) correlationism, and RQM, as we argued earlier, can be interpreted as a kind of correlationism [6]. This is consistent with the fact that there is a neo-Kantian interpretation of Rovelli's RQM, and also consistent with the interpretation of RQM ontology in terms of "metaphysical coherentism" [8–9]. Metaphysical coherence implies ontological interconnectedness (inseparability) of interacting physical systems.[4]

A similarity between RQM and agentic realism is that, from the point of view of both positions, there is no essential boundary between object and observer. Rovelli writes: "Standard quantum mechanics requires us to distinguish system from observer, but it allows us freedom in drawing the line that distinguishes the two" [10, p. 1643]. At the same time, unlike Barad, Rovelli admits that the observer is not part of the quantum mechanical description. The wave function describes the quantum system, not the observer or the quantum system and the observer. As for the so-called measurement problem, according to Rovelli, "The unitary evolution does not break down for mysterious physical quantum jumps, due to unknown effects, but simply because O is not giving a full dynamical description of the interaction. O cannot have a full description of the interaction of S with himself (O), because his information is correlation, and there is no meaning in being correlated with oneself" [10, p. 1666]. (Here "O" is a physical system playing the role of an observer and "S" is an observed physical system).

Let us consider in this connection the so called "Wigner's friend paradox".

Wigner's friend measures the magnitude of the projection of the electron spin on the z-axis (thus, the observer/observed "dualism" is assumed). Wigner observes his friend making the measurement, from outside – that is, without measuring (observing) this physical quantity. Before his measurement of the projection of electron spin on the z-axis, Wigner observes a complex system and describes it by a wave function, which is a superposition of two wave functions entangled between them: the wave function of his friend and the wave function of the observable physical system (the projection of electron spin).[5]

---

[3] Faye writes: «Bohr flatly denied the ontological thesis that the subject has any direct impact on the outcome of a measurement. Hence, when he occasionally mentioned the subjective character of quantum phenomena and the difficulties of distinguishing the object from the subject in quantum mechanics, he did not think of it as a problem confined to the observation of atoms alone. For instance, he stated that already "the theory of relativity reminds us of the subjective character of all physical phenomena". Rather, by referring to the subjective character of quantum phenomena he was expressing the epistemological thesis that all observations in physics are in fact context-dependent. There exists, according to Bohr, no view from nowhere in virtue of which quantum objects can be described» [5]. In this part, our CQR coincides with Bohr's position.

[4] In a private communication, when asked if this is not a kind of correlationism in the sense of Q. Meillassoux, M. Morganti told me that he considers a symmetric relation of physical systems, whereas Meillassoux considers a symmetric relation between the subject and the world.

From the point of view of CQR, Wigner and his friend are situated in different contexts. In relation to Wigner, the entangled system "Wigner's friend – physical system (electron)" is an external object, whereas in relation to Wigner's friend, external are Wigner and the electron he observes. By making a measurement of "the same" physical quantity that his friend measures, Wigner enters a new context, which is not necessarily the same as that of his friend. Therefore, the values of the physical quantity measured by Wigner and his friend can be different.

Rovelli speaks about "relative facts" (a classical analogue is relativity of velocity)[6], but not about facts of subject/object correlation (for him "context" means "observer", i.e. the physical system with which the "observed" physical system interacts). Relative facts are "facts" about quantum systems. Wigner and Wigner's friend are dealing with different relative facts. Therefore, there is no contradiction between the fact that O (Wigner's friend) observes some particular value of the physical quantity P pertaining to physical system S (this is a fact relative to O), and another relative fact that O' (Wigner) does not observe any particular value of P, but observes the entanglement of systems S and O (O, as stated above, cannot observe his own entanglement with S).

For our CQR, ontology, and thus the observed values of physical quantities, are context-sensitive. O and O' observe (in different contexts) different slices of reality. But what they observe is indeed real (although it is not in the "outside world", since this concept is meaningless from the CQR perspective) and independent of the observer – in this sense it is absolute rather than relative.

In relation to Wigner's paradox, Barad's position is as follows: "We are either describing a mark on the 'measuring agency' (e.g., a pointer pointing in a definite direction), in which case what it measures is its correlation with the system with which it intra-acts, constituting a particular phenomenon; or we make a different placement of the agential cut, using a different experimental arrangement such that the complete 'original' phenomenon, this time including what was previously marked as the 'measuring agency', is being measured by the 'new' 'measuring agency', in which case it is possible to characterize the existing entanglement" [2, p. 348]. Thus, for Barad, S and O always remain ontologically entangled in measurement. The problem is then that it is not clear how the particular value of the physical quantity O being measured arises. In contrast to Barad, Rovelli, as has been said, explains this (another question is how well). Barad cannot accept Rovelli's explanation without abandoning her ontological relational holism and its implications – agential realism.

In conclusion, we agree with Faye that Barad's agent realism is not a necessary consequence of quantum mechanics. Moreover, in our view, it is a misinterpretation of quantum mechanics. Faye, however, acknowledges the usefulness of agential realism for social theorizing and, accordingly, warns against the danger of using quantum mechanics for this purpose. We think that this warning is unfounded. On the contrary, agential realism, and precisely because it misinterprets quantum mechanics, is in fact also unsuitable for social theorizing, for which taking into account the sensitivity of epistemology and ontology to context is fundamental. Social reality is "quantum", not "agential", reality [11]. (See also [12–16].)

---

[5] Note that the fact that Wigner observes a complex system (superposition) consisting of his friend and the electron observed by the friend, indicates that some measurement has already been made by him. (Otherwise there would be no determinateness at all.) This is the measurement of the physical quantity for which the observed superposition is an eigenfunction.

[6] The so-called QBism does the same. However, for QBism, quantum facts are relative to the individual subjective (phenomenological) experience of the observer.